\documentclass[12pt]{iopart}
\usepackage{latexsym}
\usepackage{amsfonts}
\usepackage{epsfig}
\def\theequation{\arabic{section}.\arabic{equation}}

\renewcommand{\theequation}{\thesection.\arabic{equation}}
\begin{document}
\def\bfone{\relax{\rm 1\kern-.35em 1}}
\def\bfnull{\relax{\rm O \kern-.635em 0}}
\def\dop{{\rm d}\hskip -1pt}
\def\a{\alpha}
\def\b{\beta}
\def\g{\gamma}
\def\d{\delta}
\def\e{\epsilon}
\def\ve{\varepsilon}
\def\t{\theta}
\def\l{\lambda}
\def\m{\mu}
\def\n{\nu}
\def\pg{\pi}
\def\r{\rho}
\def\s{\sigma}
\def\t{\tau}
\def\c{\chi}
\def\p{\psi}
\def\o{\omega}
\def\G{\Gamma}
\def\D{\Delta}
\def\T{\Theta}
\def\L{\Lambda}
\def\Pg{\Pi}
\def\S{\Sigma}
\def\O{\Omega}
\def\pb{\bar{\psi}}
\def\cb{\bar{\chi}}
\def\lb{\bar{\lambda}}
\def\i{\imath}
\def\oL{\overline{L}}
\def\eq#1{(\ref{#1})}
\newcommand{\be}{\begin{equation}}
\newcommand{\ee}{\end{equation}}
\newcommand{\ba}{\begin{eqnarray}}
\newcommand{\ea}{\end{eqnarray}}
\newcommand{\ban}{\begin{eqnarray*}}
\newcommand{\ean}{\end{eqnarray*}}
\newcommand{\nn}{\nonumber}
\newcommand{\nin}{\noindent}
\newcommand{\fgl}{\mathfrak{gl}}
\newcommand{\fu}{\mathfrak{u}}
\newcommand{\fsl}{\mathfrak{sl}}
\newcommand{\fsp}{\mathfrak{sp}}
\newcommand{\fusp}{\mathfrak{usp}}
\newcommand{\fsu}{\mathfrak{su}}
\newcommand{\fp}{\mathfrak{p}}
\newcommand{\fso}{\mathfrak{so}}
\newcommand{\fg}{\mathfrak{g}}
\newcommand{\fr}{\mathfrak{r}}
\newcommand{\fe}{\mathfrak{e}}
\newcommand{\rE}{\mathrm{E}}
\newcommand{\rSp}{\mathrm{Sp}}
\newcommand{\rSO}{\mathrm{SO}}
\newcommand{\rSL}{\mathrm{SL}}
\newcommand{\rSU}{\mathrm{SU}}
\newcommand{\rUSp}{\mathrm{USp}}
\newcommand{\rU}{\mathrm{U}}
\newcommand{\rF}{\mathrm{F}}
\newcommand{\R}{\mathbb{R}}
\newcommand{\C}{\mathbb{C}}
\newcommand{\Z}{\mathbb{Z}}
\newcommand{\Hb}{\mathbb{H}}
\def\oL{\overline{L}}

%%%%%%%%%%%%%%%%%%%%%%%%%%%%%%%%%%%%%%%%%%%%%%%%%%%%%%%%%%%%%%%%%%%%%%%%%%%%%%%%%%%%%%%%%%%%%%%%%%%%%%%%%%%%%%%%%%%%%%%%%%%%%%%%%%%%%%%%%%%%
%%%%%%%%%%%%%%%%%%%%%% FINE MANIFOLDWALKER MACRO %%%%%%%%%%%%%%%%%%%%%%%%%%%%%%%%%%%%%%%%%%%%%%%%%%%%%%%%%%%%%%%%%%%%%%%%%%%%%%%%%%%%%%%%%%%
\begin{titlepage}

\begin{center}
{\LARGE {$N=4$ supergravity for Type IIB on $T^6/Z_2$\\ in presence of fluxes}}\\
\vskip 1.5cm
  {\bf  Silvia Vaul\`a} \\
\vskip 0.5cm
\end{center}
\begin{center}
{\small  Dipartimento di Fisica, Politecnico di
Torino,\\ Corso Duca degli Abruzzi 24, I-10129 Torino, Italy}\\
{\small and}\\
{\small  Istituto Nazionale di Fisica Nucleare (INFN) -
Sezione di
Torino,\\ Via P. Giuria 1, I-10125 Torino, Italy}\\
\end{center}
\begin{center}
e-mail: silvia.vaula@polito.it
\end{center}
\vskip 1cm
\begin{abstract}
We report on the construction of four dimensional gauged supergravity models that can be interpreted as type IIB orientifold compactification in presence of 3-form fluxes and $D3$--branes. We mainly address our attention to the symplectic embedding of the $U$--duality group of the theory and the consequent choice of the gauge group, whose  four dimensional killing vectors are the remnant of the ten dimensional fluxes. We briefly discuss the structure of the scalar potential arising from the gauging and  the properties of the killing vectors in order to preserve some amount of supersymmetry. 

\end{abstract}
\vfill
\begin{center}
{\small Contribution to the proceedings of the workshop of the RTN Network\\ "The quantum structure of space-time and the geometric nature of fundamental interactions"\\ Copenhagen, September 2003}
\end{center}

\end{titlepage}

\section{Introduction}

It has recently been shown that compactification of higher dimensional theories in presence of $p$--form fluxes, 
\cite{Polchinski:1995sm}, \cite{Taylor:1999ii}, \cite{Curio:2000sc}, \cite{Louis:2002ny}, \cite{Tripathy:2002qw}, \cite{Andrianopoli:2003jf}, \cite{Gukov:1999ya} can give origin to four dimensional vacua with spontaneously broken supersymmetry and vanishing cosmological constant. 
In particular \cite{Giddings:2001yu},  \cite{Frey:2002hf}, \cite{Kachru:2002he}, \cite{Berg:2003ri} orientifold compactification of type IIB supergravity in presence of 3--form fluxes leads to gauged four dimensional supergravities endowed with a positive semidefinite scalar potential. 
With respect to conventional compactifications, which typically give rise to models where most of the scalar fields are moduli, the scalar potential generated by the fluxes has a no--scale structure \cite{Ferrara:2002bt}, \cite{Cremmer:1983bf}, \cite{Ellis:1983sf} which can fix most of the scalars in the vacuum configuration where the supersymmetry can be partially or completely broken\cite{ma}--\cite{Ferrara:1988jx}.\\ 
It is interesting to understand which low energy models  correspond to such compactifications. In fact, the construction of the theory in four dimensions allows to describe also the fermionic degrees of freedom, whose derivation is quite difficult compactifying from higher dimensions, therefore it is possible to describe the Higgs and super--Higgs phases in the low energy supergravity theory.\\
The main disclaimer between all the possible four dimensional models, is the way the $U$--duality group is embedded in the symplectic group.
In the case of Type IIB superstring compactified on a
$T_6/\mathbb{Z}_2$ orientifold the relevant
embedding of the supergravity fields corresponds to the subgroup
${\rm SO}(6,6)\times {\rm SL}(2,\mathbb{R})\subset{\rm Sp}(24,\mathbb{R})$ which acts linearly on the gauge
potentials, six each coming from the NS and the RR 2--forms
$B_{\m \L}$, $C_{\m \L}$ $\L=1\dots 6$. It is obvious that
${\rm GL}(6,\mathbb{R})\subset{\rm SO}(6,6)$, which comes from the moduli space of $T^6$, does not mix the vectors coming from the RR sector with the ones coming from the NS sector, while ${\rm SL}(2,\mathbb{R})$ which comes from the Type IIB ${\rm SL}(2,\mathbb{R})$
symmetry in ten dimensions exchanges RR and NS vectors.\\ This means that the twelve vectors
are not in the fundamental {\bf 12} of ${\rm SO(6,6)}$ but rather a
$({\bf 6_+,2})$ of ${\rm GL}(6,\mathbb{R})\times {\rm SL}(2,\mathbb{R})$
where the $"+"$ refers to the ${\rm O}(1,1)$ weight of
${\rm GL}(6,\mathbb{R})={\rm O}(1,1)\times {\rm SL}(6,\mathbb{R})$. Their magnetic
dual are instead in the $({\bf 6_-,2})$ representation. Note that
instead in the heterotic string, the twelve vectors $g_{\m \L}$,
$B_{\m \L}$ are in the $({\bf 6_+^+,6_-^+})$ and
their magnetic dual in the $({\bf 6_+^-,6_-^-})$ representation,
where the lower plus or minus refer to the $\mathbb{R}$ of
${\rm GL}(6,\mathbb{R})$ and the upper
plus or minus refer to the $\mathbb{R}$ of ${\rm SL}(2,\mathbb{R})$.\\
A different choice of the symplectic embedding can lead to new gauged supergravities because inequivalent embeddings can assign the field strengths to be electric or magnetic in a different way. In fact, it was shown in a recent investigation \cite{Andrianopoli:2002mf}
\cite{Andrianopoli:2002aq}, extending previous analysis
\cite{Cremmer:1984hj}, \cite{Tsokur:1994gr}, that in order to study new forms of
$N$--extended gauged supergravities, one can look for inequivalent
maximal lower triangular subgroups of the full $U$--duality algebra
 inside the symplectic algebra of electric--magnetic
duality transformations.\\
The paper, which is based on the results of \cite{D'Auria:2002tc}, \cite{D'Auria:2002th}, \cite{D'Auria:2003jk} is organized as follows:\\
In Section 2 we describe some general aspects of the symplectic embedding and discuss the properties of the relevant one for the $T^6/\mathbb{Z}_2$ orientifold.\\
In Section 3 we illustrate the main properties of the gauged theory.\\
In section 4 we show some aspects of the supersymmetry breaking.

\section{The symplectic embedding and duality rotations}
\setcounter {equation}{0} \addtocounter{section}{0} 
Let us now briefly recall the meaning and the relevance of the symplectic embedding of the $U$--duality group  ${\rm
SL}(2,\mathbb{R})\times {\rm SO}(6,6+n)$, focusing our attention just on the bulk part ${\rm
SL}(2,\mathbb{R})\times {\rm SO}(6,6)$ \cite{Andrianopoli:1996ve}. \\
The twelve field strengths associated to the six vectors in the gravitational multiplet and the ones of the six vector multiplets in the bulk sector, come from the ten dimensional R-R and NS two forms, where an index takes values on the space-time $\m=0,1,2,3$ and the other on the internal manifold $\L=1,\dots 6$: $\{C_{MN},\,B_{MN}\}\longrightarrow\{A^1_{\m\L},\,A^2_{\m\L}\}$\\ 
We recall that $\L$ is  a ${\rm GL}(6,\mathbb R)$ index, since this latter is the structural group of the torus, and that $\{C_{MN},\,B_{MN}\}$ form an ${\rm SL}(2,\mathbb{R})$ doublet and so must do $\{A^1_{\m\L},\,A^2_{\m\L}\}$. Thus we arrange the fields of the gravitational and vector multiplets as $A_{\a\m}^{\L}$, where $\a=1,2$ is the ${\rm SL}(2,\mathbb{R})$ index, and define the electric field strengths as $F^{\L}_{\m\n\a}\equiv\partial_{[\m} A^{\L}_{\n]\a}$. The kinetic terms of the scalars and of the vector fields of the supersymmetric  ungauged Lagrangian have the following form:
\be\mathcal{L}=\label{lagvect}-2i\mathcal{N}_{\L\a\S\b}\mathcal{F}_{\m\n}^{+\L\a}\mathcal{F}^{+\S\b\m\n}+c.c.+g_{ab}\partial_{\m}\ell^{a}\partial^{\m}\ell^{b}+\frac{1}{2}G_{rr'}\partial_{\m}s^r\partial^{\m}s^{r'}\ee
Here $\mathcal{F}_{\m\n}^{\pm\L\a}$ refer to the imaginary (anti) self--dual part of the vector field strength
\be\mathcal{F}_{\m\n}^{\pm\L\a}\equiv\frac{1}{2}(\mathcal{F}_{\m\n}^{\L\a}\pm i\ ^*\mathcal{F}_{\m\n}^{\L\a});\quad ^*\mathcal{F}_{\m\n}^{\L\a}\equiv\frac{1}{2}\epsilon_{\m\n\r\s}\mathcal{F}^{\r\s\L\a}\ee
The scalars parametrize the $\s$--model $\frac{{\rm SL}(2,\mathbb{R})}{{\rm SO}(2)}\otimes\frac{{\rm SO}(6,6)}{{\rm
SO}(6)\times {\rm SO}(6)}$ and the matrices $g_{ab}$ and $G_{rr'}$ are the invariant metrics on the two factors of the coset and $\ell^a$, $s^r$ the scalar fields corresponding to the coset coordinates. \\ 
The kinetic matrix \be\mathcal{N}_{\L\a\S\b}=(\theta_{\L\a\S\b}-ig_{\L\a\S\b})\ee depends on the scalars and generalizes the coupling constant and the $\theta$--angle.\\ 
The definition of the magnetic field strengths is
\be\label{mag}\mathcal{G}^{\pm\m\n}_{\L\a}\equiv \pm\frac{i}{2}\cdot\frac{\delta\mathcal{L}}{\delta\mathcal{F}_{\m\n}^{\pm\L\a}};\quad \mathcal{G}^{+\m\n}_{\L\a}=\mathcal{N}_{\L\a\S\b}\mathcal{F}^{+\S\b\m\n};\ \mathcal{G}^{-\m\n}_{\L\a}=\overline{\mathcal{N}}_{\L\a\S\b}\mathcal{F}^{-\S\b\m\n}\ee
A transformation  of ${\rm
SL}(2,\mathbb{R})\times {\rm SO}(6,6)$, $\ell^a\longrightarrow\ell^{a\, \prime},\ s^r\longrightarrow s^{r\, \prime}$
is obviously an invariance of the scalar kinetic terms, since
\be g(\ell,\,\ell)=g(\ell^\prime\,\ell^\prime);\quad G(s,\,s)=G(s^\prime,\,s^\prime)\ee 
are the invariant measures on the two factors of the coset space.\\
The electric--magnetic field strengths carry a couple of  indices $(\L,\,\a)$ of ${\rm GL}(6,\mathbb R)\times{\rm SL}(2,\mathbb{R})\subset{\rm
SL}(2,\mathbb{R})\times {\rm SO}(6,6)$, thus they  transform linearly
\be\label{durot}\begin{pmatrix}{\mathcal{F}^{\pm\m\n\L\a}\cr \mathcal{G}^{\pm\m\n}_{\L\a} }\end{pmatrix}'=\begin{pmatrix}{A^{\L\a}_{\phantom{aA}\S\b}&B^{\L\a\S\b}\cr C_{\L\a\S\b}&D_{\L\a}^{\phantom{aA}\S\b}}\end{pmatrix}\begin{pmatrix}{\mathcal{F}^{\pm\m\n\S\b}\cr \mathcal{G}^{\pm\m\n}_{\S\b} }\end{pmatrix}\ee
In order to preserve the definition of the magnetic field strengths \eq{mag}, the kinetic matrix must transforms as
\be\label{period}\mathcal{N}'_{\L\a\S\b}(L'^{\a},s'^r)=(C_{\L\a\D\d}+D_{\L\a}^{\phantom{aA}\G\g}\mathcal{N}_{\G\g\D\d})(A^{\D\d}_{\phantom{aA}\S\b}+B^{\D\d\Pi\pi}\mathcal{N}_{\Pi\pi\S\b})^{-1}\ee
The transformation matrix must be symplectic as a consistency condition. 
\be Q\equiv\begin{pmatrix}{A&B\cr C&D}\end{pmatrix}\in {\rm SO}(6,6)\times {\rm SL}(2,\mathbb{R})\hookrightarrow {\rm Sp}(24,\mathbb{R})\ee
In fact $\mathcal{N}_{\L\a\S\b}$ is a complex symmetric matrix, and so must be $\mathcal{N}'_{\L\a\S\b}$, thus the fractional linear transformation \eq{period} must preserve complex symmetric matrices and this correspond to the request that $Q$ is a symplectic matrix.\\
A $U$-duality transformation is not a symmetry of the classical action for generic $A,\,B,\,C,\,D$. The scalar terms \eq{lagvect} are in any case invariant by construction, while the vector term \eq{lagvect} is not invariant, unless we set $B=0$. These transformations correspond to the electric subgroup which transforms electric field strengths into electric field strengths, as one can see from equation \eq{durot} setting $B=0$.\\ 
For the ungauged theory, where the $A_{\m\n}^{\L\a}$ appear in the bosonic Lagrangian only in their kinetic terms and no sources are present, the Bianchi identities and the equations of motion are clearly invariant under $U$--duality transformations, which therefore acts as a generalized electric--magnetic duality.\\
The action of the $U$--duality group on the electric--magnetic field strength depends on the way $ {\rm SO}(6,6)\times {\rm SL}(2,\mathbb{R})$ is embedded into  $ {\rm Sp}(24,\mathbb{R})$; given one choice of the embedding it is possible to obtain an inequivalent one conjugating the matrix $Q$ with an element $\mathcal{O}$\be\label{conj}\mathcal{O}\in {\rm Sp}(24,\mathbb{R})\notin{\rm SO}(6,6)\times {\rm SL}(2,\mathbb{R});\quad Q'=\mathcal{O}Q\,\mathcal{O}^{-1}\ee
This corresponds to a change of basis by means of the matrix $\mathcal{O}$ in the space of the electric--magnetic field strengths on which $Q$ acts, that is a different assignment of the role of electric or magnetic field strengths.\\
Since the $U$--duality is not a symmetry of the action, the choice of  an inequivalent embedding \eq{conj} leads to a different Lagrangian, but describes the same non perturbative theory, since we have the same equations of motion up to an $U$--duality transformation.\\
The electric subgroup of the  $U$--duality group is instead a global invariance of the action and a suitable subgroup can be promoted to a local invariance performing the gauging.  Once the gauging is performed the $U$--duality group is broken also at the level of the non perturbative theory, since the gauging fixes a certain number of field strengths to be electric and give rise to a source term for the electric field strengths, then the equations of motion and the Bianchi identities are no more invariant under exchange of electric and magnetic field strengths.\\
As a consequence  inequivalent embeddings \eq{conj} certainly correspond to inequivalent gauged theories, since they interchange the role of electric and magnetic field strengths, thus correspond to different choices of the electric subgroup.\\
It is well known that  the standard embedding \cite{Bergshoeff:1985ms}, \cite{deRoo:1985jh}  (corresponding to Heterotic $T^6$ compactification) of the isometry algebra ${\mathfrak
sl}(2,\mathbb{R})+{\mathfrak so}(6,6)$ inside ${\mathfrak
sp}(24,\mathbb{R})$,  
${\mathfrak so}(6,6)$ is diagonal, while ${\mathfrak
sl}(2,\mathbb{R})$ acts as an electric magnetic duality. One can easily see that this can not be the case for the compactification of the IIB theory on the $T^6/\mathbb{Z}_2$ orientifold. In fact, from the definition of the electric field strengths $\mathcal{F}_{\m\n\L\a}\equiv\partial_{[\m} A_{\n]\L\a}$ we see that the $ {\rm SL}(2,\mathbb{R})$ factor of the $U$--duality group is totally electric. From the definition of the magnetic field strengths \eq{mag} it is also clear that $ {\rm SL}(2,\mathbb{R})$ transforms magnetic field strengths into magnetic field strengths, thus it must be not just electric but $12\times 12$ block diagonal.\\
Furthermore the $\L$ index on the torus transform under $ {\rm GL}(6,\mathbb{R})$, that means that $ {\rm GL}(6,\mathbb{R})\subset {\rm SO}(6,6)$ must be $6\times 6$ block diagonal inside the ${\rm Sp}(24,\mathbb{R})$ matrix, since it does not mix electric and magnetic field strengths of the R--R and NS sector. \\ All these considerations suggest to decompose the ${\rm SO}(6,6)$ generators as follows:
\be\label{dec0}\fso(6,6)=\fsl(6,\R)^0+\fso(1,1)^0+(\mathbf{15}',\mathbf{1})^{+2}+(\mathbf{15},\mathbf{1})^{-2}\ee
where the superscripts refer to the ${\rm SO}(1,1)$ grading.\\
The final result is obtained by first embedding the ${\mathfrak sl}(2,\mathbb{R})$ generators and the $\fso(6,6)$ generators, according to the splitting \eq{dec0} into ${\mathfrak
sp}(24,\mathbb{R})$ in  the usual way, then performing a symplectic conjugation \eq{conj} in order to lead the symplectic matrix in the required form. The result is the following:
\begin{eqnarray}\begin{pmatrix}{\delta\mathcal{F}^{+\m\n}_{\L 1}\cr\delta \mathcal{F}^{+\m\n}_{\L 2}\cr\delta \mathcal{G}^{+\m\n}_{\L 1}\cr\delta \mathcal{G}^{+\m\n}_{\L 2}}\end{pmatrix}=\begin{pmatrix}{A+\bf 1&0&0&-t\cr 0&-A^T-\bf1&-t&0\cr0&-t'&- A^T-\bf1&0\cr -t'&0&0& A+\bf1}\end{pmatrix}\begin{pmatrix}{\mathcal{F}^{+\m\n}_{\L 1}\cr \mathcal{F}^{+\m\n}_{\L 2}\cr \mathcal{G}^{+\m\n}_{\L 1}\cr \mathcal{G}^{+\m\n}_{\L 2}}\end{pmatrix}\end{eqnarray}
 \begin{eqnarray}\begin{pmatrix}{\delta\mathcal{F}^{+\m\n}_{\L 1}\cr\delta \mathcal{F}^{+\m\n}_{\L 2}\cr \delta\mathcal{G}^{+\m\n}_{\L 1}\cr\delta \mathcal{G}^{+\m\n}_{\L 2}}\end{pmatrix}=\left( \matrix{&\matrix{\cr -{\mathfrak s}^T \cr } & \quad\matrix{\cr 0\cr}&
\cr &\matrix{\cr 0\cr \phantom{a}} &\quad\matrix{\cr \mathfrak s\cr
\phantom{a}}&}\right)
\begin{pmatrix}{\mathcal{F}^{+\m\n}_{\L 1}\cr \mathcal{F}^{+\m\n}_{\L 2}\cr \mathcal{G}^{+\m\n}_{\L 1}\cr \mathcal{G}^{+\m\n}_{\L 2}}\end{pmatrix}\end{eqnarray}
where $A$ are the ${\mathfrak gl}(6,\mathbb{R})$ generators and $t^\prime$, $t$ the generators of the tranlations $(\mathbf{15}',\mathbf{1})^{+2}$, $(\mathbf{15},\mathbf{1})^{-2}$ respectively.\\ 
The electric subalgebra is constituted by the lower triangular matrices, that is \be\label{el1}{\mathfrak sl}(2,\mathbb{R})+{\mathfrak gl}(6,\mathbb R)+{\bf t}_{15}^{\prime+2}\ee
When $n$ vector multiplets are added, we have to include the $({\bf 6},{\bf n})^{\pm 1}$ generators; nevertheless the generalization is trivial since the matter electric and magnetic field strengths transform under $\mathfrak so(1,1)$ as in the standard case and ${\mathfrak so}(n)\subset{\mathfrak so}(6,6+n)$ is totally electric, so that the electric subalgebra of ${\mathfrak sl}(2,\mathbb{R})+{\mathfrak so}(6,6+n)$ is \be\label{el2}{\mathfrak sl}(2,\mathbb{R})+{\mathfrak gl}(6,\mathbb R)+{\bf t^\prime}_{15}^{+2}+{\mathfrak so}(n)\ee
The important fact is that the conjugation matrix $\mathcal{O}$ one has to use does not preserve the structure of a symplectic matrix in the electric subgroup, i.e. with $B=0$; that means that the gaugings one can obtain from the theories described by the two different embeddings are different.  

%%%%%%%%%%%%%%%%%%%%%%%%%%%%%%%%%%%%%%%%%%%%%%%%%%%%%%%%%%%%%%%%%%%%%%%%%%%%%%%%%%%%%%%%%%%%%%%%%%%%%%%%%%%%%%%%%%%%%%%%%%%%%%%%%%%%%%%%%%%%%%%%%%%%%%%%%%%%%%%%%%%%%%%%%%%%%%%%%%%%%%%%%%%%%%%%%%%%%%%%%%%%%%%%%%%%%%%%%%%%%%%%%%%%%%%%%%%%%%%%%%%%%%%%%
\section{The gauging of the theory}
In order to understand which is the correct gauging to perform to retrieve the four dimensional theory obtained via compactification, let us  
 consider the definition of the IIB five form $\widetilde{F}_{(5)}$. Retaining just the components that survive the orientifold projection \cite{Kachru:2002he} we can derive the following expression \be\label{der0}\widetilde{F}_{\m\L\S\G\D}\equiv\partial_{\m}C_{\L\S\G\D}-\frac{1}{2}C_{\m\L}H_{\S\D\G}+\frac{1}{2}B_{\m\L}F_{\S\G\D}\ee
For constant fluxes
\be H_{\L\S\G}=\frac{\a^\prime}{2\pi}h_{\L\S\G},\quad F_{\L\S\G}=\frac{\a^\prime}{2\pi}f_{\L\S\G},\quad h_{\L\S\G},f_{\L\S\G}\in\mathbb{Z} \ee   
 satisfying the selfduality condition \cite{Frey:2002hf}
 \be *_6h^{\L\S\D}=-f^{\L\S\D};\quad\quad  *_6f^{\L\S\D}=h^{\L\S\D}\ee 
 equation \eq{der0} becomes the definition of the covariant derivative for the scalar fields  $C_{\L\S\G\D}$, identifying $\widetilde{F}_{\m\L\S\G\D}\equiv\nabla_\m C_{\L\S\G\D}$, or, more conveniently, defining the 15 scalars of the four dimensional theory as the dual of the $C_{\L\S\G\D}$ 
\be B^{\L\S}\equiv *_6C_{\D\G\Pi\O}\ee
equation \eq{der0} becomes
\be\label{der0y}\nabla_{\m}B^{\L\S}=\partial_{\m}B^{\L\S}+\frac{\a^\prime}{16\pi}f^{\L\S\G}C_{\m\G}+\frac{\a^\prime}{16\pi}h^{\L\S\G}B_{\m\G}\ee
It is clear that equation \eq{der0y} can be interpreted as the covariant derivative of $B^{\L\S}$ where the spin one fields $C_{\m\G},\, B_{\m\G}$ gauge some suitable group for which the  constant fluxes have the interpretation of killing vectors.\\ 
In order to make contact with the theory we  constructed in references \cite{D'Auria:2002tc}, \cite{D'Auria:2002th}, \cite{D'Auria:2003jk} let us define:
\be A_{\m\G}^1\! \equiv\! B_{\m\G} ,\quad A_{\m\G}^2\!\equiv\! C_{\m\G} ,\quad f^{\L\S\G}_1\!\equiv -\frac{\a^\prime}{16\pi}h^{\L\S\G};\quad f^{\L\S\G}_2\!\equiv -\frac{\a^\prime}{16\pi}f^{\L\S\G}\ee 
Since the IIB theory has an ${\rm SL}(2,\mathbb R)$ duality and the RR and NS two--form constitute an  ${\rm SL}(2,\mathbb R)$ doublet, it follows that both the four dimensional gauge fields $\{A_{\m\G}^1,\,A_{\m\G}^2\}$ and the fluxes $\{f^{\L\S\G}_1,\,f^{\L\S\G}_2\}$ are ${\rm SL}(2,\mathbb R)$ doublets, therefore we can introduce an ${\rm SL}(2,\mathbb R)$ index $\a=1,2$ and write $A_{\m\G}^\a$ and $f^{\L\S\G}_\a$. We also define, using the ${\rm SL}(2,\mathbb R)$ invariant Ricci tensor $\epsilon_{\a\b}$ \be A_{\m\G\a}\equiv\epsilon_{\a\b}  A_{\m\G}^\b,\quad f^{\L\S\G}_\a\equiv\epsilon_{\a\b}f^{\L\S\G\b}\ee     
According to the previous definitions, the covariant derivative of $B^{\L\S}$ \eq{der0y} takes the form 
\be\label{der0z}\nabla_{\m}B^{\L\S}=\partial_{\m}B^{\L\S}+f^{\L\S\G\a}A_{\m\G\a}\ee
In order to understand which is the gauge group for the four dimensional theory, one can consider the residual gauge symmetry 
\ba&&A_{\m\G\a}\longrightarrow A_{\m\G\a}+\partial_\m\eta_{\G\a}\\
&&B^{\L\S}\longrightarrow B^{\L\S}+k^{\L\S};\quad\quad k^{\L\S}\equiv f^{\L\S\G\a}\eta_{\G\a} \ea
from which it is clear that the fields $B^{\L\S}$ parametrize translational isometries on the four dimensional $\s$--model and therefore equation \eq{der0z} describes the covariant derivative with respect to a translational group which corresponds to the ${\bf t^\prime}_{15}^{+2}$ part of the decompositions \eq{el1}, \eq{el2}.\\ According to the decomposition \eq{el2}, when $D3$--branes degrees of freedom are introduced, we can also gauge a suitable group $G\subset {\rm SO}(n)$, beside 12 out of the 15 translations  ${\bf t\prime}_{15}^{+2}$.\\
The gauging of the theory modifies the structure of the Lagrangian and of the supersymmetry transformation laws of the fermions. The latter acquire extra terms, whose properties strongly depend on the choice of the gauging, in the present case they have the following structure
\ba\label{G1}
\d\p_{A\m}&=&-\frac{i}{48}G^-_{AB}\g_\m\e^B;\quad\quad\d\c^A=-\frac{1}{48}G^{+\,AB}\e_B\\
\d\l^{A(\overline{4})}&=&\frac{1}{8}G^-_{AB}\e_B;\quad\quad\quad\,\ \label{G4}\d\l_A^{I\,\,(\overline{20})}=\frac{1}{48}\overline{G}^{\,-}_{AC}(\G^{I})^{BC}\e_B\\
\d\l_{iA}&=&W^{\,\,\,B}_{iA}\e_B\label{G5}
\ea 
In equations \eq{G1}, \eq{G4} the 24 dimensional representation of ${\rm
SU}(4)_{(d)}\subset {\rm SU}(4)_1\times {\rm SU}(4)_2$ to which the bulk gaugini
$\l_A^I$ belong has been decomposed into its irreducible parts, namely
$\bf{24}=\bf{\overline{20}}+\bf{\overline{4}}$. Setting:
\be\l_A^I=\!\l_A^{I\,\,(\overline{20})}-\frac{1}{6}(\G^{I})_{AB}\l^{B(\overline{4})};\quad\l^{A(\overline{4})}=\!(\G_{I})^{AB}\l_B^I;\quad(\G_{I})^{AB}\l_B^{I\,\,(\overline{20})}=0\ee
The matrices appearing in equations \eq{G1}, \eq{G4} are defined as follows
\ba &&G_{AB}^-=(\overline{F}^{IJK-}+\overline{C}^{IJK-})(\G_{IJK})_{AB};\quad \overline{G_{AB}^-}=\overline{G}^{AB+}\nn\\
&&G^{AB+}=(\overline{F}^{IJK+}+\overline{C}^{IJK+})(\G_{IJK})^{AB};\quad \overline{G^{AB+}}=\overline{G}_{AB}^-\ea  
The quantities  $C^{IJK\pm}$, $F^{IJK\pm}$, $W^{\,\,\,B}_{iA}$ that  depend on the scalars and on the killing vectors of the gauge group are defined in reference  \cite{D'Auria:2003jk}.\\
The gauging modifies also the Lagrangian, in particular gives rise to a scalar potential, that on general ground  is a linear combination of the traced moduli square of the ${\rm SU}(4)$ matrices appearing in the fermion shifts \eq{G1}, \eq{G5}, where all the coefficients are positive but the one of the gravitino contribution. Note that the gravitino shift matrix and the $\overline{\bf 4}$ gaugino shift matrix have the same structure $G_{AB}^-$, while the shift matrix of the  dilatino is the complex conjugate of the ${\bf 20}$ gaugino one. It turns out that the contribution to the potential of the gravitino and the  ${\bf 4}$ gaugino cancel against each other, therefore the potential is positive semidefinite and depends just on the moduli square of the shift matrix of the dilatino and the ${\bf 20}$ gaugino, plus the contribution of the $D3$--brane sector gaugini:  
\be V=\frac{1}{(12)^2}|G^{+\,AB}|^2+{W_{iB}}^A{W^{iB}}_A\ee
If we consider just the bulk sector, where the matter gaugini $\l^i_A$ are absent and $C^{IJK\pm}=W^{\,\,\,B}_{iA}=0$, the matrices $G^{\pm}_{AB}$ are the four dimensional remnant of the ten dimensional flux $G$ of references \cite{Grana:2001xn}, \cite{Frey:2002hf} and the potential reduces to 
\be  V=\frac{1}{(12)^2}|G^{+\,AB}|^2\ee
which vanishes for imaginary antiselfdual fluxes $G^{+\,AB}=0$ according to reference \cite{Kachru:2002he}. (Note that we define \cite{D'Auria:2002tc}, \cite{D'Auria:2002th}, \cite{D'Auria:2003jk} the imaginary (anti)selfdual projections as $F^{\pm}=\mp i*F^{\pm}$ and we call ``imaginary self dual'' the projection $F^+$, while in reference  \cite{Kachru:2002he} the opposite convention is used).   
\section{The supersymmetry breaking}
The minimum of the potential is reached when  $G^{+\,AB}=W^{\,\,\,B}_{iA}=0$; the condition that sets  $W^{\,\,\,B}_{iA}=0$ also implies that  $C^{IJK\pm}=0$ and then the remaining condition is $\overline{F}^{IJK+}=0$ or its complex conjugate $F^{IJK-}=0$. The condition  $W^{\,\,\,B}_{iA}=0$ sets to zero all the $D3$--brane sector scalars except the ones transforming in the Cartan subalgebra of the gauge group $G$; note that this implies that the contribution of the $D3$--brane sector to the supersymmetry transformation laws in the vacuum is identically vanishing and therefore just the translational gauging is responsible of the supersymmetry breaking that involves only the bulk sector degrees of freedom. The condition $F^{IJK-}=0$ is satisfied for a fixed value of the complex dilaton, say $\Phi=i$, and when the fluxes satisfy the selfduality condition \be f_1^{\L\S\G}=*f^{\L\S\G}_2\ee that explicitly reads as
\begin{equation}\label{dual}
f_1^{\L\S\G}=\frac {1}{3!}{\rm det}g^{-\frac{1}{2}}
\e^{\L\S\G\D\Pi\O}g_{\D\D'}g_{\Pi
\Pi'}g_{\O\O'}f_2^{\D'\Pi'\O'}
\end{equation}
where $g^{\L\S}\equiv E^{\L}_I E^{\S }_J\eta^{IJ}$ is the (inverse) moduli metric of $T^6$ and the vielbein $ E^{\L}_I$ intertwines between the ${\rm GL}(6,\mathbb{R})$ indices $\L,\,\S,\,\G$ and the ${\rm SO}(6)$ indices $I,\,J,\,K$. \\
Note that, beside the trivial case where all the fluxes are set to zero,  even if $F^{IJK-}=0$ we can not set also  $F^{IJK+}=0$, therefore the presence of fluxes breaks supersymetry as one can see from equations \eq{G1}--\eq{G4}. In order to discuss the supersymmetry breaking is useful to introduce a complex basis starting from  the basis vectors $\{e_\L\}$, $(\L=1\dots 6)$ of
the fundamental representation of ${\rm GL}(6,\mathbb{R})$ and defining a complex basis $\{E_i,\overline{E}_i\}$ with
 $i=x,\,y,\,z$ in
the following way:
\begin{eqnarray}\label{complex} &&e_1+ie_4
=E_x;\,\,\,e_2+ie_5=E_y;\,\,\,e_3+ie_6=E_z
\\&&e_1-ie_4
=\overline{E}_x;\,\,\,e_2-ie_5=\overline{E_y};\,\,\,e_3-ie_6=\overline{E_z}
\end{eqnarray}
We are now able to analyze the condition on the metric moduli coming from equation \eq{dual}.  Going to the
complex basis and lowering the indices on both sides, one obtains an equation
relating a $(p,q)$ form on the l.h.s $(p,q=0,1,2,3;\,p+q=3)$ to a
combination of $(p',q')$ forms on the r.h.s. Requiring that all
the terms with $p'\neq p,\,q'\neq q$ are zero and that the r.h.s.
of equation \eq{dual} be a constant, on is led to fix different
subsets of the $g^{\L\S}$ moduli, depending on the
number of translational symmetries that is gauged.\\ 
Furthermore the diagonalized gravitino shift matrix is given by
\be G^+_{AB}=-i\oL^2\,\left[f_1^{-\overline{x}\overline{y}\overline{z}}\G_{ \overline{x}\overline{y}\overline{z}}+f_1^{-x\overline{y}z}\G_{x\overline{y}z}+f_1^{-\overline{x}yz}\G_{\overline{x}yz}+f_1^{-xy\overline{z}}\G_{xy\overline{z}}\right]_{AB}\label{snoopy}\ee
where one can see that only the $(0,3)$ and primitive $(2,1)$ components of the fluxes appear. The shift matrix \eq{snoopy} is proportional to  the gravitino mass matrix, therefore from its eigenvalues one can read the gravitino masses
\begin{eqnarray}\label{zeromass}
&&m_1\equiv|\m_1+i\m_1^{'}|=\frac{1}{6|L^2|}|f^{-\overline{x}yz}_1|;\quad\quad m_2\equiv|\m_2+i\m_2^{'}|=\frac{1}{6|L^2|}|f^{-xy\overline{z}}_1|\\
&&m_3\equiv|\m_3+i\m_3^{'}|=\frac{1}{6|L^2|}|f^{-x\overline{y}z}_1|;\quad\quad m_4\equiv|\m_4+i\m_4^{'}|=\frac{1}{6|L^2|}|f^{-\overline{x}\overline{y}\overline{z}}_1|\label{zeromassx}
\end{eqnarray}
Since the fluxes are independent we can set an arbitrary number of masses equal to zero and show \cite{D'Auria:2002th}, \cite{D'Auria:2002tc}, \cite{D'Auria:2003jk} that it is possible to realize the supersymmetry breaking to $N=3,2,1,0$. If we do not want to break supersymmetry completely we must set at least one gravitino mass to zero. Note that it is irrelevant which mass we set to zero because it just depends on the choice of the complex structure on the torus; one natural choice could be to set $m_4=0$, retrieving the results of \cite{Grana:2001xn}, that is that the fluxes must be primitive and of type $(2,1)$ in order to preserve some amount of supersymmetry. Nevertheless with a change of complex structure, for instance $(E_x,\,E_y,\,E_z)\rightarrow(\overline{E}_x,\,\overline{E}_y,\,E_z)$ the corresponding choice would be $m_2=0$.

\section*{Acknowledgements}

This report is based on collaborations with R. D'Auria, S. Ferrara, F. Gargiulo, M. A. Lled\'o and M. Trigiante, that  the author would like to thank.\\  Work
supported in part by the European Community's Human Potential
Program under contract HPRN-CT-2000-00131 Quantum Space-Time, in
which the author is associated to Torino University.

\end{document}